\documentclass[aps,superscriptaddress,twocolumn]{revtex4-1}

\usepackage{subfigure,dcolumn}
\usepackage{epstopdf}
\usepackage{mhchem}
\usepackage{upgreek}
\usepackage{float}
\usepackage{graphicx}
\usepackage{color}
\usepackage[T1]{fontenc}

\usepackage{xcolor}
\usepackage{ulem}

\begin{document}

\title{Light and hyper nuclei formation at $\sqrt{s_{\text{NN}}} =$ 3 GeV Au+Au collisions using  Wigner coalescence approach}

\author{L.~K. Liu}
\affiliation{Key Laboratory of Quark {\rm \&} Lepton Physics (MOE) and Institute of Particle Physics, Central China Normal University, Wuhan 430079, China}
\author{C.~L Hu}
\affiliation{University of Chinese Academy of Sciences, Beijing 100049, China}
\author{X.~H. He}
\affiliation{Institute of Modern Physics, Chinese Academy of Sciences, Lanzhou 730000, China}
\author{S.~S. Shi}
\affiliation{Key Laboratory of Quark {\rm \&} Lepton Physics (MOE) and Institute of Particle Physics, Central China Normal University, Wuhan 430079, China}
\author{G.~N. Xie}
\affiliation{University of Chinese Academy of Sciences, Beijing 100049, China}

\begin{abstract}
The production of light nuclei and hyper-nuclei in heavy-ion collisions, particularly at high baryon density, is crucial for understanding the dynamical evolution of the collision system and exploring the internal state of nuclear matter of compacted stellar object. Despite being a topic of ongoing debate, an improved theoretical understanding is necessary. 
In this work, production of light nuclei ($d$, $t$, $^{3}$He, $^{4}$He) and hyper-nuclei ($^{3}_{\Lambda}$H, $^{4}_{\Lambda}$H) was investigated using the JAM microscopic transport model combined with an afterburner coalescence process at $\sqrt{s_{\text{NN}}} =$ 3 GeV Au+Au collisions. The formation of a specific nucleus during the coalescence process is determined by its Wigner function.
The comparison of the calculations for $\mathrm{p_T}$ spectra, average $\mathrm{p_T}$, and rapidity distributions to the measurements from the STAR experiment was performed.
We investigated the dynamic information carried by light nuclei and determined the averaged spatial distance $\langle \Delta R \rangle$ and momentum difference $\langle \Delta P \rangle$ of constituent nucleons ($\Lambda$) for each nucleus species.
\end{abstract}


\maketitle

\section{Introduction}
\label{introduction}
The neutron star is one of the most densest objects in the universe, with central densities reaching several times of the density of normal nuclear matter. The understanding of the internal composition of neutron stars, which is still an open question, depends on the equation of state (EOS) of strongly interacting matter. The present way to reproduce this extreme condition in laboratory is through high-energy heavy-ion collisions, where the two colliding nuclei are stopped to create high baryon density nuclear matter.
Moreover, it is expected that hyperons will appear in the core of neutron stars as their weak decay is suppressed by the Pauli blocking of the phase space available for nucleons~\cite{VA_1960, PhysRevLett.114.092301, PhysRevC.81.035803}. 
The EOS of neutron stars is strongly influenced by the presence of hyperons and the fundamental hyperon-nucleon (YN) interactions, which affects their mass and radius~\cite{PhysRevC.81.035803, PhysRevC.84.035801, Massot_2012}. Hyper-nuclei, a bound state of hyperon and nucleons, provides a laboratory for studying YN interactions~\cite{RevModPhys.88.035004, Ma_2023}. In particular, the formation of hyper-nuclei in the dense nuclear matter created in heavy-ion collisions may provide direct insight into the YN interactions.

However, the production mechanisms of both light nuclei and hyper-nuclei in heavy-ion collisions are still under debate. Given that their binding energy is on the order of MeV, significantly lower than the temperature of the collision system, which is on the order of hundreds of MeV, it is likely that they can only be produced at the later stages of the system's evolution when the temperature is lower, through the coalescence of hyperon and nucleons. In the coalescence picture, the nucleons (hyperon) will combine when they are close to each other in both the coordinate and momentum spaces~\cite{SATO1981153, PhysRevC.99.014901,Mattiello:1996gq, SUN2015272}. 
The probability of coalescence is determined by the spatial distance $\Delta R$ and momentum difference $\Delta P$ within the constituent baryons. For different light nuclei or hyper-nuclei species, these parameters could be different reflecting their intrinsic characteristics. 
Additionally, for a hyper-nucleus with the same atomic mass number as a light nucleus, differences in their $\Delta R$ and $\Delta P$ could also arise from distinctions in nucleon-nucleon (NN) interactions and YN interactions.
The recent measurements of the $\mathrm{p_T}$ spectrum and yield of light nuclei and hyper-nuclei~\cite{2023arXiv231111020T,PhysRevLett.128.202301} encourage us to investigate the possible difference in the formation of different nuclei species.

In this work, we explore the light nuclei ($p$, $d$, $t$, $^{3}$He, $^{4}$He) and hyper-nuclei ($^{3}_{\Lambda}$H, $^{4}_{\Lambda}$H) formation in Au+Au collisions at $\sqrt{s_{\text{NN}}} =$ 3 GeV using the coalescence approach of Wigner function.
The paper is organized as follows: In Sec. II, we introduced the coalescence method and Wigner functions employed for each studied nuclei species. 
In Sec. III, the calculations for the $\mathrm{p_T}$ spectra, $\langle \mathrm{p_T} \rangle$, 
and $dN/dy$ are compared to the data. 
The ratio of coalescing nucleons to free ones is determined in terms of the transverse radius $\mathrm{r_T}$ and transverse momentum $\mathrm{p_T}$.
The average relative distances $\langle \Delta R \rangle$ and relative momenta $\langle \Delta P \rangle$ between constituent nucleons are extracted for each nucleus species.
In Sec IV, we provide a summary of this study.

\section{method}
The coalescence model assumes that during the late stage of system evolution, 
nucleons close to each other in phase space combine together to form light- or hyper-nuclear clusters. 
As a result, the yield of light- and hyper-nuclei is determined by the phase space distribution of nucleons and the correlations among them. In this study, the final phase space distribution of nucleons ($\Lambda$) is populated using the newly developed Jet AA Microscopic Transport Model (JAM)~\cite{PhysRevC.105.014911,PhysRevC.106.044902}.

JAM is a microscopic transport model developed to investigate the collision dynamics of high-energy heavy-ion collisions. 
The initial positions of nucleons are sampled by the distribution of nuclear density, 
and subsequent collisions are described by the sum of independent nucleon-nucleon collisions. There are two different transport modes: cascade mode and mean-field mode. In the cascade mode, the trajectory of each hadron is assumed to be straight until it encounters a collision with other hadrons. 
In the mean-field mode, the hadron feels an additional potential arising from the surrounding nuclear matter. At $\sqrt{s_{\text{NN}}} =$ 3 GeV Au+Au collisions, several experimental measurements, including particle yields and collective flow, indicate that the mean-field potential plays a crucial role, and its calculations can qualitatively reproduce the observed data~\cite{STAR:2021fge, STAR:2022etb, 2023arXiv231111020T, STAR:2021yiu, STAR:2021ozh, Lan:2022rrc}. 
Thus, the mean-field mode of JAM is employed to produce the phase distribution of nucleons and hyperons, where their positions and momenta are recorded at a fixed evolution time of 20 fm/$c$ for further coalescence calculations. The choice of using 20 fm/$c$ for the coalescence time is based on the fact that calculations at this time provide an optimal description of the data compared to the results obtained with a shorter or longer time. 

To form light nuclei ($p$, $d$, $t$, $^{3}$He, $^{4}$He) and hyper-nuclei ($^{3}_{\Lambda}$H, $^{4}_{\Lambda}$H), the coalescence of the generated nucleons and $\Lambda$ is carried out using Wigner functions.
The Wigner function represents the probability of a single particle in the spatial momentum space, which can be expressed as:
\begin{equation}
w(\mathbf{r},\mathbf{p}) = \int d^3y\left<\mathbf{r}+\mathbf{y} / 2|\Psi\right>\left<\Psi| \mathbf{r}-\mathbf{y} / 2\right>\exp(-\frac{i}{\hbar} \mathbf{p} \cdot \mathbf{y})
\end{equation}
where $\Psi(\mathbf{r})$ is the wave function of the particle, 
$\mathbf{r}$ is the position, $\mathbf{p}$ is the momentum, and $\mathbf{y}$ represents a small displacement.
Neglecting the hopefully small effect from binding energies, the probability of producing a nucleus is determined by the overlap of its Wigner phase–space density with the nucleon phase–space distribution near kinetic freeze-out~\cite{Mattiello:1996gq,PhysRevC.102.044912,PhysRevC.98.054905,ZHAO2021136571} .

The Wigner density function of deuteron, denoted as $\rho_2^{W}$ and analogues to the single particle Wigner function, can be derived as follows:
\begin{equation}
\rho_2^W(\mathbf{r}, \mathbf{k})=\int d^3y \Psi\left(\mathbf{r}+\frac{\mathbf{y}}{2}\right) \Psi^{\ast}\left(\mathbf{r}-\frac{\mathbf{y}}{2}\right) 
\exp(-i \mathbf{k} \cdot \mathbf{y})
\label{Eq_wigner2}
\end{equation}
where $\mathbf{k}=\left(m_2 \mathbf{p}_1-m_1 \mathbf{p}_2\right) /\left(m_1+m_2\right)$ and $\mathbf{r}=\left(\mathbf{r}_1- \mathbf{r}_2 \right)$ are the relative momentum and relative coordinate, respectively, and $\Psi(\mathbf{r})$ is the relative wave function of the two constituents. 
The wave function of deuteron can be represented using the wave function of a spherical harmonic oscillator~\cite{PhysRevC.59.1585,Ko:2012lhi}
\begin{equation}
\Psi(\mathbf{r})=\left( \pi \sigma^2_{\mathrm{d}}\right)^{-3 / 4} \exp\left(-\frac{r^2}{2\sigma^2_{\mathrm{d}}}\right)
\label{Eq_DWaveFunc}
\end{equation}
The parameter $\sigma_{\mathrm{d}} $ is associated with the root-mean-square (rms) radius of deuteron, given by
$R^2=\int d^3 r\left(\frac{r}{2}\right)^2\left|\Psi(\mathbf{r})\right|^2$, with a value of 3.2 fm.
Following Eq.~\ref{Eq_wigner2}, one finds the $\rho_2^W(\mathbf{r}, \mathbf{k})$ is
\begin{equation}
\rho_2^W(\mathbf{r}, \mathbf{k}) = 8 \exp \left(-\frac{r^2}{\sigma_{\mathrm{d}}^2}-\sigma_{\mathrm{d}}^2 k^2\right)
\label{Eq_Wigner2}
\end{equation}
where $\mathbf{r}$ and $\mathbf{k}$ are the relative coordinate and the relative momentum of the proton and neutron in the deuteron rest frame.

For triton and ${}^3 \mathrm{He}$, solving the three-body problem exactly is challenging, 
and people usually take approximations to simplify the problem. 
One of the most popular and effective approaches is the hyper-spherical method, in which the wave function of triton and ${}^3 \mathrm{He}$ is defined to be similar to that of the deuteron. 
Their Wigner phase-space densities are derived from their internal wave functions, which are assumed to follow the pattern of a spherical harmonic oscillator~\cite{PhysRevC.98.054905,ZHAO2021136571,PhysRevC.59.1585,Ko:2012lhi,Chen:2003ava, Zhang:2021vsf},

\begin{equation}
\Psi\left(\mathbf{\rho},\mathbf{\lambda}\right)=\left(3 \pi^2 \sigma^4\right)^{-3 / 4} \exp \left(-\frac{\rho^2+\lambda^2}{2 \sigma^2}\right)
\end{equation}
The $\mathbf{\rho}$ and $\mathbf{\lambda}$ are the relative coordinates of in the rest frame of triton and ${}^3 \mathrm{He}$, respectively. 
They are obtained through the transformation of the three-nucleon space position $\mathbf{r}_1, \mathbf{r}_2, \mathbf{r}_3$ using the normal Jacobi matrix $J$. Similarly, the relative momenta $\mathbf{k_\rho}$ and $\mathbf{k_\lambda}$ are defined from  $\mathbf{k}_1, \mathbf{k}_2, \mathbf{k}_3$.

\begin{equation}
\left(\begin{array}{c}
\mathbf{R} \\
\mathbf{\rho} \\
\mathbf{\lambda}
\end{array}\right)=J\left(\begin{array}{l}
\mathbf{r}_1 \\
\mathbf{r}_2 \\
\mathbf{r}_3
\end{array}\right)
\quad
\left(\begin{array}{c}
\mathbf{K} \\
\mathbf{k}_\rho \\
\mathbf{k}_\lambda
\end{array}\right)=(J^{-})^T\left(\begin{array}{l}
\mathbf{k}_1 \\
\mathbf{k}_2 \\
\mathbf{k}_3
\end{array}\right)
\end{equation}

\begin{equation}
J=\left(\begin{array}{ccc}
\frac{1}{3} & \frac{1}{3} & \frac{1}{3} \\
\frac{1}{\sqrt{2}} & -\frac{1}{\sqrt{2}} & 0 \\
\frac{1}{\sqrt{6}} & \frac{1}{\sqrt{6}} & -\frac{2}{\sqrt{6}}
\end{array}\right)
\end{equation}
The Wigner phase-space densities for triton and ${ }^3 \mathrm{He}$ is then given by:
\begin{equation}
\rho_3^W(\mathbf{\rho}, \mathbf{\lambda}, \mathbf{k_\rho}, \mathbf{k_\lambda}) = 8^2 \exp \left(-\frac{\mathrm{\rho}^2+\mathrm{\lambda}^2}{\sigma^2} -\left(k_\rho^2+k_\lambda^2\right) \sigma^2\right)
\label{eq_r3}
\end{equation}
The root-mean-square radius $R$ of triton or ${ }^3 \mathrm{He}$ can be written as:
\begin{equation}
R^2=\int \frac{\rho^{2}+\lambda^2} {2}\left|\Psi\left(\mathbf{\rho}, \mathbf{\lambda}\right)\right|^2 3^{3 / 2} d \rho d \lambda= \sigma^2
\end{equation}
The parameter $\sigma$ is determined to be 1.59 fm and 1.76 fm for triton and ${ }^3 \mathrm{He}$, respectively, 
based on their measured rms radius~\cite{Ropke:2008qk}.

By using the hyper-spherical method, the 4th order Jacobi matrix with determinant $|J| = 1 / \sqrt{4}$ is utilized to 
describe the transformation of the relative momentum and relative coordinate for the 4-body ($^{4}$He) Wigner density function,
which can be expressed as:
\begin{equation}
\rho_{4}  = 8^3 \exp \left(-\frac{\rho^{2} + \lambda^{2} +\gamma^2} {\sigma^{2} }  
 -(k_{\rho}^2 +k_{\lambda}^2 + k_{\gamma}^2 ) \sigma^2 \right)
 \label{eq_r4}
\end{equation}
where $\mathbf{\rho}$, $\mathbf{\lambda}$, $\mathbf{\gamma}$ and $\mathbf{k_\rho}$, $\mathbf{k_\lambda}$, $\mathbf{k_\gamma}$ are the three relative coordinate and relative momentum parameters in the 4-body rest frame, 
the parameter $\sigma$ is $\sqrt{ \frac{8}{9}} $ times of the nuclei radius~\cite{SUN2015272}.

For a $\Lambda$ hyper-nucleus, the total binding energy is significantly larger than the $\Lambda$ separation energy.
Therefore, hyper-nuclei are proposed to be produced through the coalescence of $\Lambda$ and light nuclei core, such as $d+\Lambda\rightarrow ^{3}_{\Lambda}$H, and $t+\Lambda\rightarrow ^{4}_{\Lambda}$H.
Consequently, the Wigner density distribution of hyper-nuclei takes a 2-body form similar to that of the deuteron~\cite{Zhang:2018euf}.

Table ~\ref{tab1} summarizes the parameters used to determine the Wigner density function for each light nucleus and hyper-nucleus species in this study~\cite{Ropke:2008qk, Sun:2015ulc, Zhang:2018euf}. The spin factor $g_c$ is calculated as $g_{c} = (2j+1)/2^{N}$, where $j$ is the spin of the nucleus, and $N$ is the number of constituent baryons. The column for binding energy, with brackets for hyper-nuclei, represents the binding energy of the inner light nuclei and the $\Lambda$ separation energy.
The spin factor of ${ }_{\Lambda}^4 \mathrm{H}$ is taking into account the contribution from the excited $1^{+}$ state~\cite{BEDJIDIAN1976467,1979252,PhysRevLett.115.222501}.

\begin{table}[htbp]
    \caption{Parameters used for the Wigner density function of light nuclei and hyper-nuclei. The binding energy of hyper-nuclei is the sum of the binding energy of light nuclei core and the $\Lambda$ separation energy in the brackets~\cite{Ropke:2008qk, Sun:2015ulc, Zhang:2018euf, BEDJIDIAN1976467,1979252,PhysRevLett.115.222501}.}
\begin{tabular}{c|c|c|c}
\hline Nuclei & rms-radius $[\mathrm{fm}]$ & spin factor $g_c$ & \begin{tabular}{c} 
binding \\
energy $[\mathrm{MeV}]$ \\
\end{tabular} \\
\hline d & 1.96 & $3 / 4$ & 2.225 \\
\hline t & 1.59 & $1 / 4$ & 8.482 \\
\hline${ }^3 \mathrm{He}$ & 1.76 & $1 / 4$ & 7.718 \\
\hline${}^4 \mathrm{He}$ & 1.45 & $1 / 16$ & 28.3 \\
\hline${ }_{\Lambda}^3 \mathrm{H}$ & 4.9 & $1 / 4$ & \begin{tabular}{c}
2.354 \\
$(2.225+0.13)$
\end{tabular} \\
\hline${ }_{\Lambda}^4 \mathrm{H}$ & 2.0 & $1 / 4$ & \begin{tabular}{c}
10.601 \\
$(8.482+2.12)$
\end{tabular} \\
\hline
\end{tabular}
    \label{tab1}
\end{table}

For each event, all possible combinations of nucleons ($\Lambda$)—including 2-body, 3-body, and 4-body configurations—are boosted to their respective rest frames. The Wigner density factor is then calculated for each configuration, representing the production probability.

\section{Results and discussions}
\label{Sec:Result}

\begin{figure}[htbp]
  \centering
  \includegraphics[width=0.9\columnwidth]{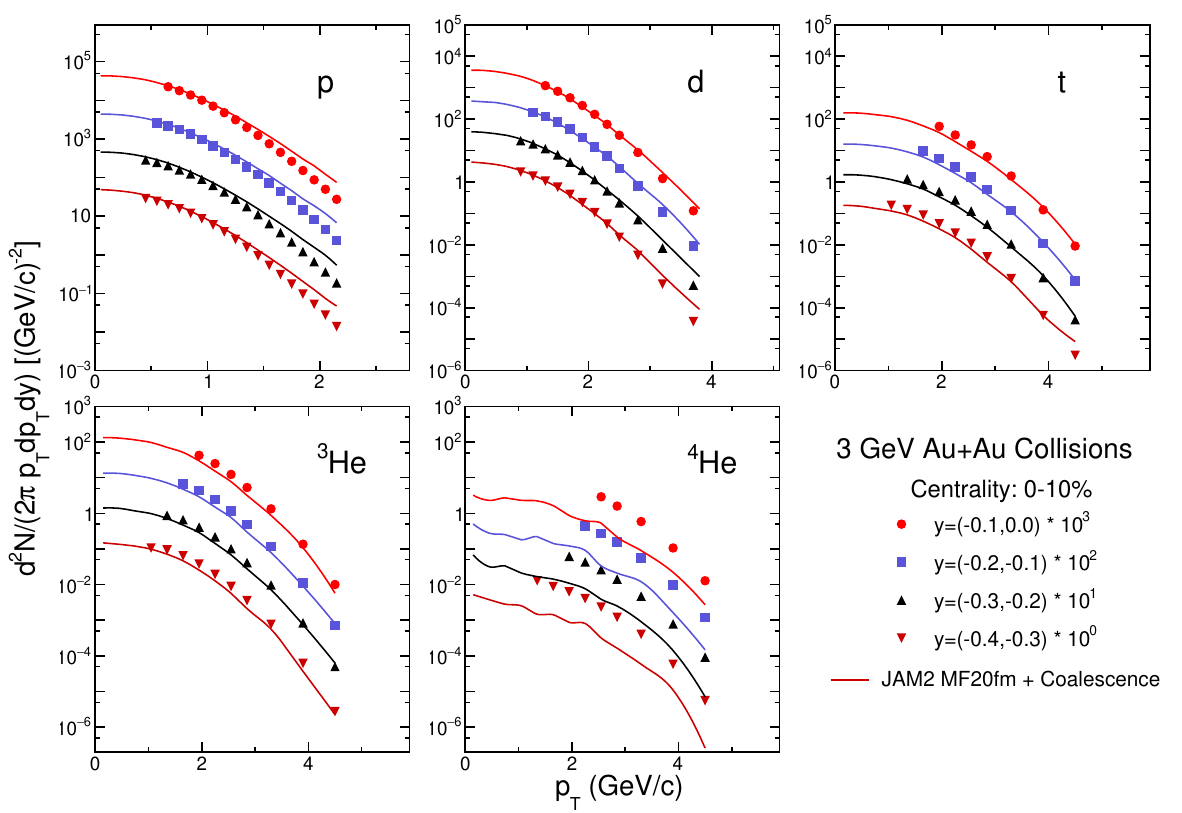}
  \includegraphics[width=0.68\columnwidth]{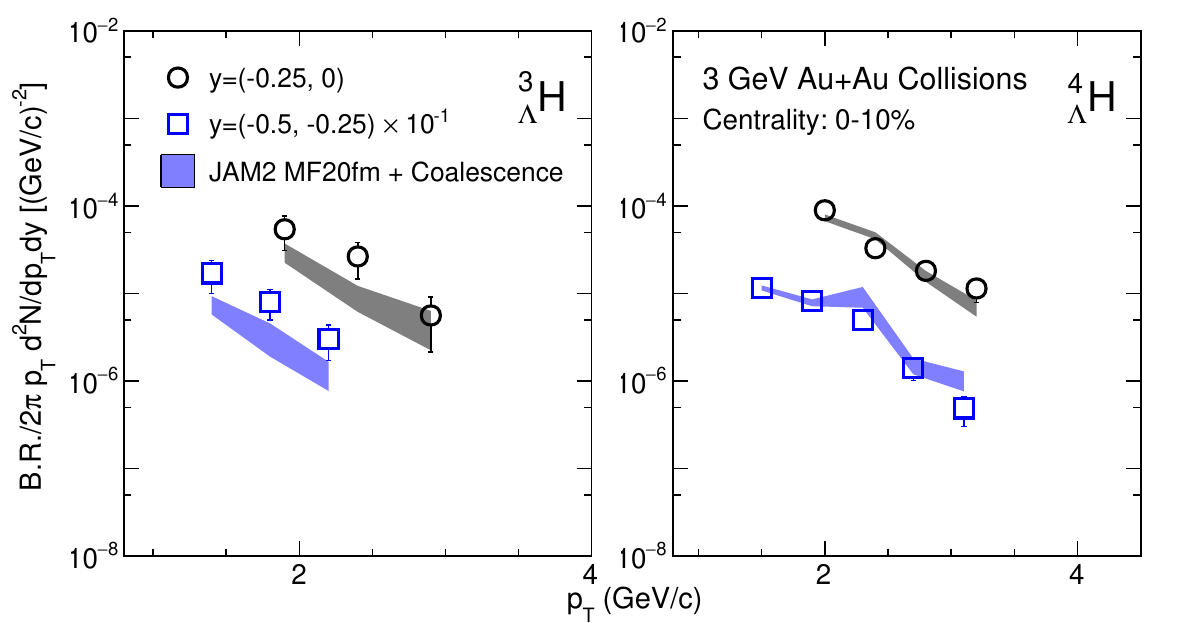}
  \caption{Transverse momentum spectra of protons, deuterons, tritons, $^{3}$He, $^{4}$He, $^{3}_{\Lambda}$H, and $^{4}_{\Lambda}$H from various rapidity ranges in the most central Au+Au collisions at $\sqrt{s_{\text{NN}}} =$ 3 GeV. Data points represent the results measured by the STAR collaboration~\cite{2023arXiv231111020T,PhysRevLett.128.202301}.
  The solid lines (bands) are the calculations of the JAM mean-field for proton and of JAM plus afterburner coalescence for light- and hyper-nuclei in this study.}     
  \label{Fig:PtSpec}
\end{figure}

Figure~\ref{Fig:PtSpec} shows transverse momentum ($\mathrm{p_T}$) spectra of proton, deuteron, triton, $^{3}$He, $^{4}$He, $^{3}_{\Lambda}$H, and $^{4}_{\Lambda}$H in various rapidity ranges in the most central Au+Au collisions at $\sqrt{s_{\text{NN}}} =$ 3 GeV. 
Data points represent the results measured by the STAR collaboration~\cite{2023arXiv231111020T,PhysRevLett.128.202301}.
The solid lines (bands) are the calculations with JAM mean-field mode over an evolution time of 20 fm/$c$, accompanied by an afterburner coalescence of nucleons ($\Lambda$).
In general, these calculations can qualitatively describe all light- and hyper-nuclei $\mathrm{p_T}$ spectra at different rapidity windows, while $^4\mathrm{He}$ stands out as an exception.
The $^4\mathrm{He}$ yield is notably underestimated by a factor of 6.
The coalescence model assumes that nucleons coalesce during the final stage of hadron gas interaction evolution as light (hyper-) nuclei with relatively small binding energies. 
However, $^4\mathrm{He}$ possesses a binding energy of 28.3 MeV, approximately 13 times that of a deuteron.
Consequently, $^4\mathrm{He}$ may form earlier than other light nuclei species or survive longer in such a hot environment which was first pointed out in paper~\cite{SUN2015272}.
A similar conclusion is drawn from calculations involving the kinetic production of light nuclei~\cite{Wang:2023gta, Cheng:2023rer}, indicating that the $^{4}$He formed is less dissolved during the evolution of the collision system.

\begin{figure}[!htb]
    \centering
    \includegraphics[width=0.5\textwidth]{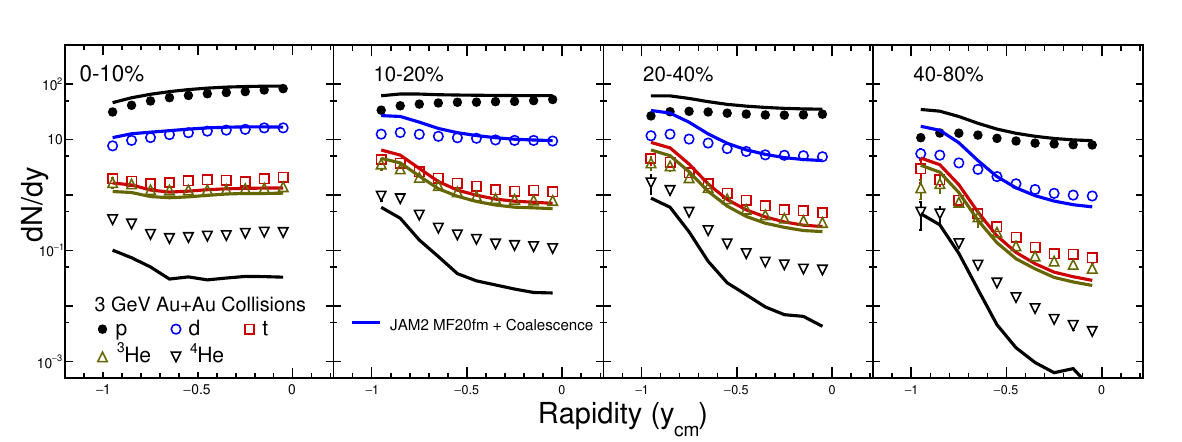}
    \caption{Protons and light nuclei $dN/dy$ in (from left to right) 0-10\%, 10-20\%, 20-40\%, and 40-80\% centrality in  $\sqrt{s_{\text{NN}}} =$ 3 GeV Au+Au collisions.  Data points represent the results measured by the STAR collaboration~\cite{2023arXiv231111020T}.
     The solid lines are the calculations of JAM mean-field for proton and of JAM plus afterburner coalescence for light- and hyper-nuclei in this study.}
    \label{Fig:dNdy}
\end{figure}

The $\mathrm{p_T}$ integrated yield as a function of particle rapidity ($dN/dy$) for light nuclei in various centrality intervals is illustrated in Fig.~\ref{Fig:dNdy}. The calculations qualitatively reproduce the data across centrality bins for deuteron, triton, and $^{3}$He, while they consistently underestimate the $dN/dy$ of $^{4}\mathrm{He}$ across all rapidity ranges and centrality bins.
Furthermore, an overestimation of the light nuclei yield is observed near the target rapidity ($y_{\text{cm}}\sim-1$), particularly in more peripheral collisions. This discrepancy can be attributed to the contribution of spectator fragmentation in the initial collisions, a process expected to be significant near the target rapidity. 
The JAM model does not account for fragment production in its calculations. The nucleons not involved in the initial collisions are transported independently. Consequently, the proton yield near the target rapidity is overestimated in the model, leading to an overestimation of the light nuclei yield in the coalescence process.

\begin{figure}[htb]
    \centering
    \includegraphics[width=0.47\textwidth]{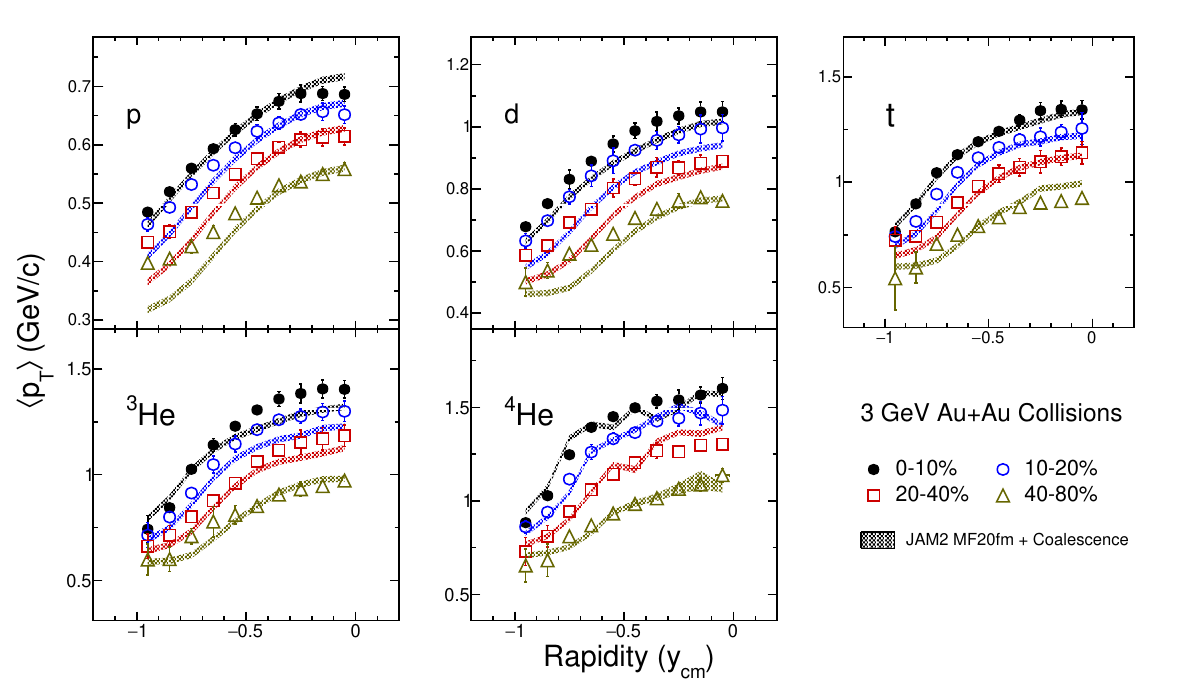}
    \caption{ Mean transverse momentum ($\langle \mathrm{p_T} \rangle$) as a function of particle rapidity for protons and light nuclei in different centrality intervals at $\sqrt{s_{\text{NN}}} =$ 3 GeV Au+Au collisions. Data points represent the results measured by the STAR collaboration~\cite{2023arXiv231111020T}.
    The dash bands are the calculations of JAM mean-field for proton and of JAM plus afterburner coalescence for light nuclei.}
    \label{Fig:meanpT}
\end{figure}

Figure~\ref{Fig:meanpT} represents the mean transverse momentum ($\langle \mathrm{p_T} \rangle$) as a function of particle rapidity across different centrality intervals at $\sqrt{s_{\text{NN}}} =$ 3 GeV Au+Au collisions. 
The JAM calculations generally reproduce the proton $\langle \mathrm{p_T} \rangle$ at the mid-rapidity 
but underestimate it near the target rapidity, especially in the peripheral collisions. 
Similar to proton $dN/dy$, this may be caused by the lack of treatment for spectator fragmentation in the JAM calculations. 
Our coalescence calculations also qualitatively describe the data for all studied light nuclei species.
However, unlike the $\mathrm{p_T}$ spectra and $dN/dy$, the $\langle \mathrm{p_T} \rangle$ of $^{4}$He can be well reproduced by the calculations, which means the coalescence assumption can reproduce the shape of $\mathrm{p_T}$ spectra. 
As this shape is determined by the collisions before the kinetic freeze-out of the evolution, this observation suggests that $^{4}$He may have formed via nucleon coalescence, but at an early stage after the kinetic freeze-out compared to deuteron, triton, and $^{3}$He.

\begin{figure}[htbp]
  \centering
  \includegraphics[width=1.0\columnwidth]{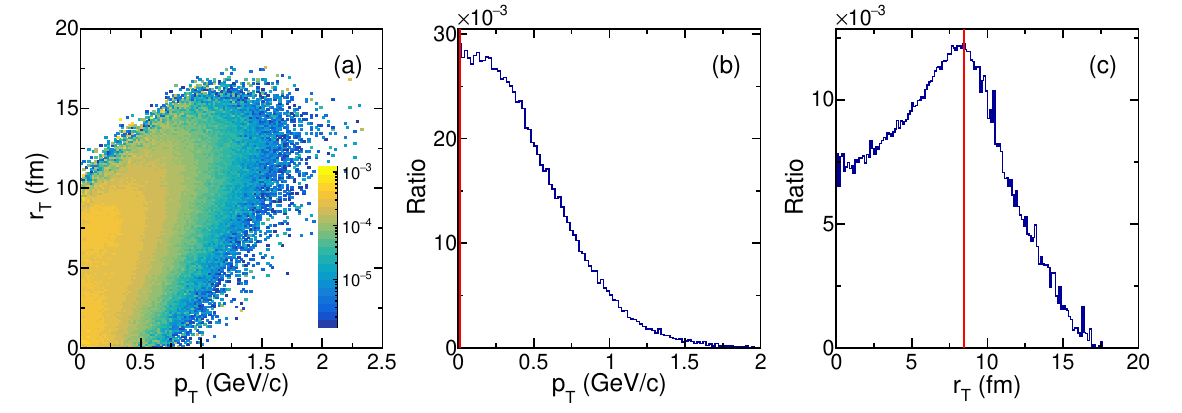}
  \caption{(a) Ratios of nucleons participated in the coalescence over that of free nucleons at kinetic freeze-out
  as a dependence of transverse momentum ($\mathrm{p_T}$) and transverse distance ($\mathrm{r_T}$).
  (b) The projection of ratios on transverse momentum. (c) The projection of ratios on transverse distance.}     
  \label{Fig:Ratio_freeCoal}
 \end{figure}

 In high-energy nuclear collision, freeze-out is a dynamical process that depends not only on the interaction cross-section of each particle but also on the overall expansion rate. If light nuclei are produced dominantly via coalescence, one may ask: Any additional dynamic information can be learned from the measurements of light nuclei? To answer this question, in the simulation we separate those freeze-out nucleons that are used in the coalescence of deuteron from those that are free from the process. The result is shown by the ratio of the coalescence nucleon over that of free ones in transverse radius $\mathrm{r_T}$ and transverse momentum $\mathrm{p_T}$, see plot (a) in Fig.~\ref{Fig:Ratio_freeCoal}. The projections to $\mathrm{p_T}$ and $\mathrm{r_T}$ are shown in plots (b) and (c), respectively. As one can see, those nucleons who participated in the coalescence are predominantly from lower $\mathrm{p_T}$ and near the surface of the fireball. In other words, one may treat coalescence as an additional freeze-out condition for those nucleons that are part of the light nuclei. On average, they carry additional information that cannot be extracted from those free nucleons. This is the case for deuteron production and the same must be true for the cases of other light nuclei.

In a simplified coalescence model, the relative spatial distances $\Delta R$ and relative momenta $\Delta P$ between the constituent nucleons serve as the criteria for forming light nuclei. In the calculations, these two parameters are varied for different light nuclei species~\cite{Steinheimer:2012tb, Li:2016mqd, Sombun:2018yqh}. 
Therefore, they are expected to be associated with the intrinsic properties and unique characteristics of different light- and hyper-nuclei. We extracted the $\Delta R$ and $\Delta P$ in the rest frame of the formed nuclei based on the equation below:

\begin{equation}
\begin{aligned}
\Delta R_{2} & = r_{12} \\
\Delta R_{3} & = \sqrt{ \rho^2 + \lambda^2} \\&= \sqrt{\frac{1}{3}\left(r_{12}^2+r_{13}^2+r_{23}^2\right)}  \\
\Delta R_{4} & = \sqrt{ {\rho^{2} + \lambda^{2} +\gamma^2} } \\&= \sqrt{\frac{1}{4}\left(r_{12}^2+r_{13}^2+r_{14}^2 + r_{23}^2 + r_{24}^2 +r_{34}^{2} \right)} 
\end{aligned}
\end{equation} 
where $\Delta R_{2}$, $\Delta R_{3}$ and $\Delta R_{4}$ represents relative distance for 2-components, 3-components and 4-components, respectively.
The $r_{ij}$ is the relative distance between two particles $i$ and $j$ in their rest frame. The $\mathbf{\rho}$,  $\mathbf{\lambda}$, and $\mathbf{\gamma}$ are defined in Eq.~\ref{eq_r3} and Eq.~\ref{eq_r4}. The corresponding $\Delta P$ is defined using the same method.

\begin{figure}[htb]
    \centering
    \includegraphics[width=0.47\textwidth]{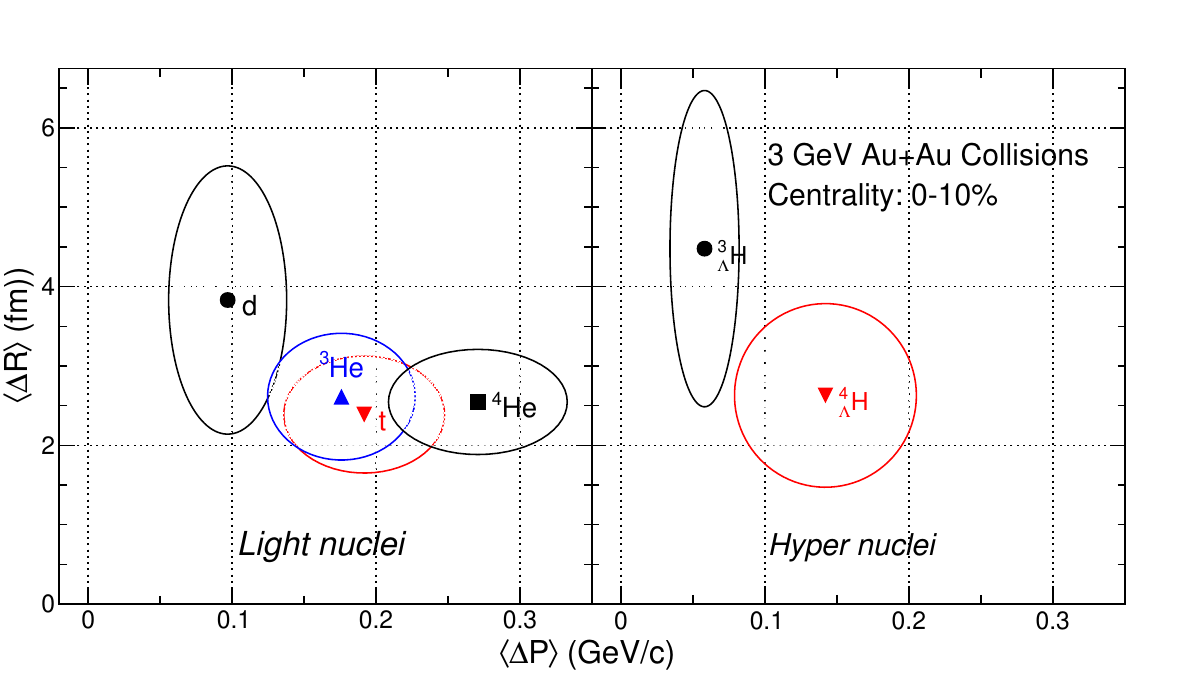}
    \caption{The average values of the relative distance $\Delta R$ and relative momentum $\Delta P$ between constituent nucleons in light- and hyper-nuclei extracted in the Wigner coalescence calculation. In the left panel, the solid circle, down triangle, upper triangle, and square represent the results for deuteron, triton, $^{3}$He, and $^{4}$He, respectively. In the right panel, the solid circle and down triangle represent the results for $ ^{3}_{\Lambda}$H and $ ^{4}_{\Lambda}$H. 
    The standard deviation of $\Delta R$ and $\Delta P$ are represented by ellipses.
    }
    \label{Fig:dRdP}
\end{figure}

Figure~\ref{Fig:dRdP} shows the average relative distance $\Delta R$ and relative momentum $\Delta P$ between constituent nucleons for light nuclei and between the light nuclei core and $\Lambda$ for hyper-nuclei. 
As the atomic mass number increases, $\Delta R$ for the light nuclei decreases, while $\Delta P$ increases. This trend is consistent for $^{3}_{\Lambda}$H and $^{4}_{\Lambda}$H as well.
The results suggest that nuclei with larger radii exhibit larger $\Delta R$ in the coalescence process, while nuclei with higher binding energy show larger $\Delta P$ values. 
This implies that nuclei with tighter binding, which are more resistant to breaking, 
allow constituent nucleons to carry a greater momentum difference. 
The average value of $\langle \Delta R \rangle$ for $ ^{3}_{\Lambda}$H is notably higher than those for tritons and $^{3}$He, whereas the opposite trend is observed for $\langle \Delta P \rangle$. This discrepancy can be attributed to the fact that the $ ^{3}_{\Lambda}$H is a loosely bound system. These observations hold across different centrality bins.

\section{Summary}
The production of light nuclei ($d$, $t$, $^{3}$He, $^{4}$He) and hyper-nuclei ($^{3}_{\Lambda}$H, $^{4}_{\Lambda}$H) was studied using the JAM microscopic transport model combined with an afterburner coalescence in $\sqrt{s_{\text{NN}}} =$ 3 GeV Au+Au collisions.
The phase space distribution of nucleons and $\Lambda$ was generated by the JAM calculation with the mean-field mode, and their positions and momenta were recorded at a fixed time of 20 fm/$c$ for subsequent coalescence calculations. The production probability in the coalescence process was determined by the Wigner function of each nucleus.
Our calculation can qualitatively reproduce the data of $\mathrm{p_T}$ spectra, average $\mathrm{p_T}$, and $dN/dy$ distributions measured by the STAR experiment for all studied nuclei species, except for $^{4}$He.
The ratio of coalescing nucleons to free ones is calculated in terms of the transverse radius $\mathrm{r_T}$ and transverse momentum $\mathrm{p_T}$, suggesting that deuterons or, in general, light nuclei formed via the coalescence process probe a slightly different fireball region compared to free nucleons.
The spatial distance $\Delta R$ and momentum difference $\Delta P$ of constituent nucleons ($\Lambda$) were extracted for each nucleus species. It was observed that $\Delta R$ decreases with increasing atomic mass number for both light nuclei and hyper-nuclei, while $\Delta P$ increases.
However, the yield of $^{4}\mathrm{He}$ was underestimated by a factor of 6, possibly due to its large binding energy, which could lead to its early formation time.


\section*{Acknowledgments}
We are grateful for discussions with Drs. Yasushi Nara and Nu Xu. This work is supported in part by the National Natural Science Foundation of China under Grant No. 12205342, 12175084 and the Strategic Priority Research Program of Chinese Academy of Sciences under Grant No. XDB34000000.

\bibliographystyle{woc.bst}
\bibliography{ref}

\end{document}